\documentclass[prd,11pt, twocolumn,
 reprint,
nofootinbib,
 amsmath,amssymb,
 aps,
prd]{revtex4-2}

\usepackage{graphicx}
\usepackage{dcolumn}
\usepackage{bm}
\usepackage[hidelinks, colorlinks = true, linkcolor = blue, urlcolor = blue, citecolor = red]{hyperref}

\usepackage{tabularray}
\usepackage{lipsum}
\usepackage{comment}
\usepackage{cancel}
\usepackage{slashed}
\usepackage{float}
\usepackage{bbold}
\usepackage{hyperref}
\hypersetup{
    colorlinks=true,
    linkcolor=blue, 
    citecolor=blue,
    urlcolor=blue
}

\usepackage{xcolor} 
\definecolor{oceanboatblue}{rgb}{0.0, 0.47, 0.75}
 
\definecolor{mygreen}{RGB}{0,128,0}

\begin{document}

\title{Scalar Dark Matter Production through the Bubble Expansion Mechanism: \\ The Role of the Lorentz factor and Non-Renormalizable Interactions}

\author{Jose A. R. Cembranos}
\email{cembra@ucm.es}
\author{Jesús Luque}
\email{jesluque@ucm.es}
\author{Javier Rubio}
\email{javier.rubio@ucm.es}
\affiliation{Departamento de Física Teórica and Instituto de Física de Partículas y del Cosmos (IPARCOS-UCM), Facultad de Ciencias Físicas, Universidad Complutense de Madrid, 28040 
Madrid, Spain}

\begin{abstract}
We consider a \textit{Bubble Expansion} mechanism for the production of scalar dark matter during a first-order phase transition in the very early Universe. Seeking for a dark matter energy density in agreement with observations, we study different renormalizable and non-renormalizable interactions between the dark matter species and the field undergoing the transition, considering all possible regimes for the Lorentz boost factor associated with the motion of the bubble wall. By employing a combination of analytical and numerical techniques, we demonstrate that sufficient dark matter production is achievable even in the previously unexplored low-velocity bubble expansion regime, enlarging the parameter space and possibilities of the scenario. Notably, for the non-renormalizable interactions it is found that the produced dark matter abundances exhibit a similar qualitative behavior to the renormalizable case, even for low Lorentz boost factors. Furthermore, for a transition around the electroweak scale, the associated gravitational wave spectrum is within the reach of future detectors.
\end{abstract}       

\maketitle


\section{\label{sec:intro}INTRODUCTION}

Despite the extensive gravitational evidence for Dark Matter (DM)—such as precise measurements of the Cosmic Microwave Background, anomalies in the rotation curves of spiral galaxies and different gravitational lensing effects—, its true nature remains a mystery, with numerous experimental searches  yielding so far no results \cite{Cirelli:2024ssz}. 

Particle DM production mechanisms can be broadly classified into thermal and non-thermal processes, each of them with distinct characteristics and implications. Thermal mechanisms, such as \textit{Freeze-Out} \cite{FreezeOUt, GONDOLO1991145, GriestandSeckel}, involve DM particles that were once in thermal equilibrium with the Standard Model plasma but ``freezed out" as the Universe expanded, leading behind a relic density generically depending on the annihilation cross-section. This results in a predictable relationship between the mass and the interaction strength of DM particles, being the former quantity limited to at most 100 TeV by unitarity considerations \cite{Griest:1989wd,Baldes:2017gzw,Smirnov:2019ngs}. In contrast, non-thermal mechanisms, such as \textit{Freeze-In} \cite{McDonald:2001vt,Choi:2005vq,Kusenko:2006rh,Petraki:2007gq,FIHall:2009bx, FIBernal:2017kxu,Elahi:2014fsa,Bernal:2020bfj,Bernal:2020qyu}, produce DM particles through extremely weak interactions, preventing them from reaching thermal equilibrium with the Standard Model plasma. This translates into an inverse relationship between the relic DM abundance and the interaction strength, allowing effectively for a broader range of possible masses and couplings. Alternatively, First-Order Phase Transitions (FOPTs) could play a role in the production of DM relics, either in the form of dark monopoles \cite{Murayama:2009nj,Khoze:2014woa,Bai:2020ttp} or of particle DM \cite{Falkowski:2012fb,Freese:2023fcr,Mansour:2023fwj,Shakya:2023kjf,Giudice:2024tcp,Baldes:2023fsp,Baldes:2024wuz}. 

In the context of FOPTs, the so-called \textit{Bubble Expansion} (BE) mechanism \cite{Bodeker:2017cim,Azatov:2020ufh,Azatov:2021ifm} allows for the copious production of very heavy DM candidates usually suppressed by Boltzmann statistics. This scenario relies on the expansion of the true vacuum bubbles generated during the FOPT and  the associated breaking of momentum conservation.  Interestingly enough, the expansion and collision of bubbles, together with the sound waves propagating in the plasma, gives rise to a sizable gravitational wave (GW) signal \cite{PTHindmarsh:2020hop, Caprini:General, SteinhartBubbles} potentially within the reach of future terrestrial and space-based experiments, such as the Einstein Telescope (ET) \cite{ET}, the Laser Interferometer Space Antenna (LISA) \cite{LISA} or the Atomic Experiment for DM and Gravity Exploration in space (AEDGE) \cite{AEDGE:2019nxb}. 

Several realizations of the BE scenario have been considered in the literature, primarily within the ultra-relativistic BE regime. These involve, for instance, the generation of scalar DM candidates via renormalizable interactions \cite{Baldes:2022oev} and the production of fermionic and vector species through Proca-like couplings \cite{Ai:2024ikj} and non-renormalizable interactions \cite{Azatov:2024crd}. In this work, we investigate the non-thermal production of scalar DM particles in BE scenarios, focusing on the so far unexplored low-velocity regime and extending the analysis to non-renormalizable interactions not previously considered for the production of scalar DM. Our findings demonstrate that a sufficient DM abundance can be achieved even in the low-velocity regime, broadening the scope and viability of the BE mechanism. Furthermore, we show that the production through non-renormalizable interactions results in no significant qualitative differences in the outcomes, apart from the expected rescaling factors associated with the number of vacuum expectation value (vev) insertions. This conclusion can be extended qualitatively even at low Lorentz boost factors, where the differences are bounded up to one order of magnitude.

This paper is organized as follows. The general scenario is introduced in Section \ref{sec:THEORETICAL BACKGROUND}, where we describe qualitatively the wall dynamics and review the essential ingredients of the BE mechanism. Our main results are presented in Section \ref{sec:BE MECHANISM}, where we determine the DM relic abundance for different renormalizable and non-renormalizable couplings between the scalar DM candidate and the FOPT sector. The associated GW signals and their potential detectability by future GW missions are discussed in Section \ref{sec: GW SIGNAL}. Finally, our conclusions are presented in Section \ref{sec:CONCLUSIONS}. 

\section{THE MODEL}\label{sec:THEORETICAL BACKGROUND}

To explore the BE mechanism in FOPT with renormalizable and non-renormalizable interactions, we consider an effective Lagrangian density
\begin{equation}
    \mathcal{L}=\partial_{\mu} \Phi\partial^{\mu}\Phi^{\dagger}+\frac{(\partial_{\mu} \phi)^2}{2}-\frac{M^2\phi^2}{2}-\frac{\lambda\Phi^n\phi^2}{n\Lambda^{n-2}}-V(\Phi)\,,
    \label{lagrangiano genérico}
\end{equation}
describing the interaction between the scalar field $\Phi$ experiencing the transition and another scalar field $\phi$ playing the role of DM, with $\lambda$ a dimensionless coupling constant, $\Lambda$ a cutoff scale signaling the onset of new physics and $V(\Phi)$ an unspecified potential allowing for the transition. The DM field $\phi$ is taken to be initially in its ground state, with the $Z_2$ symmetry $\phi\leftrightarrow -\phi$ preventing the decay of the associated DM particles upon production.

The FOPT is taken to begin when the number of bubbles per Hubble volume equals one, which,  due to the expansion and cooling of the Universe, occurs effectively when its temperature becomes lower than a critical \textit{nucleation temperature}, $T_{\text{n}}$. From this moment on, expanding bubbles appear and the scalar field $\Phi$ acquires a non-zero vev $\langle \Phi \rangle$ inside them, inducing with it an effective interaction 
\begin{equation}\label{Lint}
{\cal L}_{\rm int}=\frac{\lambda \langle \Phi\rangle^{n-1}}{\Lambda^{n-2}}\, h\,\phi^2\,,
\end{equation}
with $h(\mathbf{x})\equiv \Phi(\mathbf{x})-\langle \Phi \rangle$ the excitations of the field $\Phi$ around $\langle \Phi \rangle$. As explicit in this expression, the number of background field insertions $\langle \Phi \rangle$ entering the effective coupling  $\lambda_{\rm eff}\equiv \lambda \langle \Phi\rangle^{n-1}/\Lambda^{n-2}$ increases with the order of the operator. Note that the interaction presented in Eq.\,(\ref{Lint}) corresponds to the leading-order term obtained from Eq.\,(\ref{lagrangiano genérico}) when substituting $\Phi(x)=h(x)+\langle\Phi\rangle$. Additional interactions, leading to different final states, also emerge from this expansion. Nonetheless, we will focus on the low-velocity BE regime, where these other interactions remain subdominant. 

\subsection{WALL DYNAMICS}\label{sec:walldynamics}

The behavior of bubbles in FOPTs is a widely treated and highly complex subject \cite{CapriniPT:2015zlo,Ellis:1809,PTHindmarsh:2020hop, Caprini:General,Vanvlasselaer:2022fqf, Breitbach:2018kma, Ai:2023suz}. For the purposes of this work, however, it will be enough to describe them in terms of their characteristic scales.

The amount of energy $\epsilon$ released during the transition will be characterized by the so-called {\it transition strength parameter} $\alpha\equiv\epsilon/\rho$, with $\rho$ the energy density in the symmetric phase. In addition, we will consider the commonly named {\it transition rate parameter} $\beta$ 
measuring the rate of the transition in units of the Hubble rate $H$ and the \textit{final temperature} reached after it,
\begin{equation}
T_{\text{f}}=(1+\alpha)^{1/4}T_{\rm n}\,,
\end{equation}
which, for supercooling scenarios \cite{CapriniPT:2015zlo, Ellis:2019oqb}, can potentially exceed the nucleation temperature $T_{\rm n}$, leading with it to strong FOPTs where bubbles expand at high velocity. All these parameters, together with the Lorentz boost factor $\gamma_w$, can be determined once the characteristic potential $V(\Phi)$ governing the FOPT is specified. In this sense, the corresponding wall velocity $v_w=\sqrt{1-1/\gamma_w^2}$ could take a value close to the speed of light in the absence of friction (\textit{runaway regime}), or reach just a certain terminal value, relativistic or not, if the friction is sufficiently large to slow down the bubble wall (\textit{terminal velocity regime}) \cite{CapriniPT:2015zlo,GWCaprini:2019egz}.

\subsection{BE mechanism}\label{BE mechanism}

Given the macroscopic character of the bubbles, we consider a planar wall that expands along the negative $z$ direction and assume that the variation of the vev within it is linear.  We adopt a coordinate system in which the expanding bubble wall is at rest and the expectation value of $\Phi$ varies smoothly along the z-axis within the wall. This corresponds to
$\langle \Phi\rangle (z)=vz/L_w$,  with $v$ the vev in the non-symmetric phase ($z>L_w$), $L_w\sim 1/v$ the width of the wall and $0\leq z\leq L_w$. In the symmetric phase ($z<0$) the vev remains zero. 

In the above frame, the $h$ particles themselves approach the bubble at wall velocity $v_w$ in the $z$-direction. The breaking of translational symmetry and the associated lack of momentum conservation in the $z$-direction opens up the decay channel $h\rightarrow \phi\phi$, allowing for the exotic production of heavy DM particles with masses $M\gg v, T_{\rm n}$ \cite{Azatov:2020ufh}, a process customarily forbidden in Lorentz-invariant backgrounds. Additionally, this hierarchy of scales enables us to neglect thermal and field-dependent corrections to the bare mass parameter $M$. If the momentum of $h$ particles is also large enough, this approximation can also be extended to the symmetry-breaking sector, allowing us to treat the $h$ field as essentially massless.

The transition probability of a $h$ particle into a final state with two $\phi$ particles is given by
\begin{equation}
    P\equiv \int \frac{\textit{d}^3k^{\phi_1}\textit{d}^3k^{\phi_2}}{(2\pi)^6 2k_0^{\phi_1}2k_0^{\phi_2}} \frac{|\langle h|\mathcal{T}|\phi_1\phi_2\rangle|^2}{\langle h|h\rangle}\,,
    \label{probabilidad general}
\end{equation}
with $\mathbf{k}^{\phi_1},\mathbf{k}^{\phi_2}$ the momenta of the produced particles, $\langle h|h\rangle=1$, $\mathcal{T}$ the non-trivial contribution to the $S$-matrix, $S=\mathbb{1}+i \mathcal{T}$, and \cite{Bodeker:1703}
\begin{equation}
    \langle h|\mathcal{T}|\phi_1\phi_2\rangle=\int\mathrm{d}x^4\langle h|\mathcal{H}_{int}|\phi_1\phi_2\rangle=(2\pi)^3\delta^2_{\perp}\delta_0\mathcal{M}\,,
\end{equation}
with $\delta_{\perp}^2\equiv\delta^2(p^{h}_{\perp}-k^{\phi_1}_{\perp}-k^{\phi_2}_{\perp})$, $\delta_0\equiv\delta(p^{h}_0-k^{\phi_1}_{0}-k^{\phi_2}_{0})$ and $\mathbf{p}$ the momentum of the decaying particle and where the subscript $_{\perp}$ signals the component of the momentum orthogonal to the motion of the incoming particle $h$, i.e. orthogonal to $z$.  The matrix element $\cal M$ in this expression can be written as
\begin{equation}
    \mathcal{M}=\int \textit{d}z\chi_{\phi_1}^*(z)\chi_{\phi_2}^*(z)\chi_h(z)V(z)=\int \textit{d}ze^{i\Delta p_zz}V(z)\,, \label{matriz M}
\end{equation}
with $\chi(z)=\exp(ip_zz)$ a solution of the free particle evolution equation in the presence of the bubble wall \cite{Bodeker:1703},
\begin{equation}
    V(z)=\left\lbrace\begin{array}{cr}
        0\,, & \text{if}\quad z<0  \\
        \frac{\lambda}{\Lambda^{n-2}} \left(\frac{v\, z}{L_w}\right)^{n-1}, & \text{if} \quad0<z<L_w \\
        \lambda \frac{v^{n-1}}{\Lambda^{n-2}} \,, & \text{if} \quad z>L_w
    \end{array}\right. \,,
    \label{vertice}
\end{equation}

\noindent and $\Delta p_z=p_z-k^{\phi_1}_z- k^{\phi_2}_z$ the difference between the initial momentum of the $h$ particle and that of its decay products $\phi_1,\phi_2$ in the direction orthogonal to the wall. Note that the $(z/L_w)^{n-1}$ dependence induced by the vev insertions can also be seen as a reparametrization of the bubble wall. Thus, the presented results, apart from rescaling factors $v^{n-2}/\Lambda^{n-2}$, hold for renormalizable scenarios (n=2) with different wall ansatzes. In this case, no additional processes beyond the 2-body decay need to be considered in any velocity regime.

The value of the matrix element ${\cal M}$ can be obtained by taking into account the kinematics in the discussed frame, namely
\begin{equation}
    \begin{split}
    p & =(p_0,0,0, p_0)\,,\\
    k^{\phi_1} & =(p_0(1-x),0,k_{\perp},\sqrt{p_0^2(1-x)^2-k_{\perp}^2-M^2})\,,\\
    k^{\phi_2} & =(p_0x,0,-k_{\perp},\sqrt{p_0^2x^2-k_{\perp}^2-M^2})\,,
\end{split}
\label{momentos}
\end{equation}
with $x \in [0,1]$ guaranteeing energy conservation.\footnote{ Note that, since for the decay $h\rightarrow\phi\phi$ to occur it is necessary that $p_0\geq2M$ while $m_h\propto v$, we can safely approximate $p^h_z=\sqrt{p_0^2-m_h^2}\simeq p_0$ as long as  $M\gg v$.} 
In the considered kinematic regime, Eq.~(\ref{probabilidad general}) reduces to \cite{Bodeker:1703}
\begin{equation}
P= \frac{1}{2p_z}
         ~\int \frac{\textit{d}^3\textit{k}^{\phi_1} \textit{d}^3\textit{k}^{\phi_2}}{(2\pi)^62k^{\phi_1}_02k^{\phi_2}_0}(2\pi)^3\delta^2_{\perp}\delta_0 \big|\mathcal{M}\big|^2\,, 
        \label{probabilidad}
\end{equation} 
which, assuming $h$ to be in thermal equilibrium with the SM bath at temperature $T_n$, translates into a non-thermal number density \cite{Azatov:2021ifm}
\begin{equation}
    n_{\phi}\simeq \frac{2}{\gamma_w v_w}\int \frac{\textit{d}^3p}{(2\pi)^3} \, \mathcal{P} \, f_h(p_z,\vec{p}_{\perp}) \,,
    \label{expresion general densidad}
\end{equation}
being $\mathcal{P}\equiv \frac{p_z}{p_0}\textit{P}$ and with 
\begin{equation}
f_h(p_z,\vec{p}_{\perp})=\exp\left(-\frac{\gamma_w(E_h-v_w p_z)}{T_{\text{n}}}\right)
\label{distribución}
\end{equation}
a Boltzmann distribution, $\gamma_w\equiv 1/\sqrt{1-v_w^2}$ the Lorentz factor and $E_h=\sqrt{p_z^2+\vec{p}_{\perp}^2}$ the energy of the decaying particle $h$. In the absence of additional dilution mechanisms, the associated present DM abundance $\Omega_{\text{BE}}$ can be trivially obtained by properly redshifting the normalized DM energy density $n_\phi M$ at the time of production. Assuming the FOPT to happen during radiation domination, we have

\begin{equation}
\begin{split}
     \Omega_{\text{BE}} h^2=h^2\frac{n_{\phi} M}{\rho_c}\frac{g_{\star s0}}{g_{\star s}(T_{\text{f}})}  \left(\frac{T_0}{T_{\text{f}}}\right)^3=\Upsilon\frac{n_{\phi}M}{\text{GeV}}\frac{1}{g_{\star s}T_f^3}\,, 
\end{split}
    \label{parámetro densidad}
\end{equation}
where we have defined a numerical factor $\Upsilon=g_{\star0}T_0^3h^2\text{GeV}/\rho_c\simeq6.3\cdot10^8$, being $\rho_c=3H_0^2/8\pi G$ the critical energy density today, $T_0$ the current temperature of the Universe and $g_{\star S0} / g_{\star S}(T_{\text{f}})$ the number of entropic degrees of freedom today/at the final temperature $T_f$ \cite{Husdal:2016haj}. Note that, in this occasion, $h$ is the dimensionless Hubble parameter.

Before delving into specific cases to analyze the DM production resulting from this mechanism, it is useful to outline some general considerations regarding the computation of the decay probability and the associated DM density. As previously stated, the probability of decay can be derived from the matrix element $\cal M$ using Eq.~(\ref{probabilidad}). Notably, since $k_{\perp}^{\phi_1}=-k_{\perp}^{\phi_1}\equiv k_{\perp}$ and the dependence on $k_{\perp}$ is always quadratic, we can perform the integration over $\mathrm{d}^2k_{\perp}^{\phi_2}$ using the associated Dirac delta function. By transforming $\mathrm{d}^3k^{\phi_1}$ into cylindrical coordinates, we obtain
\begin{equation}
    \mathcal{P}=\frac{1}{32p_0\pi^2}\int\frac{\mathrm{d}k_z^{\phi_1}\mathrm{d}k_z^{\phi_2}|k_{\perp}|\mathrm{d}k_{\perp}^2}{p_0^2x(1-x)}\delta_0|\mathcal{M}|^2.
\end{equation}
Performing now an additional change of variables to $x,k_{\perp}^2,k_0^{\phi_1}$, with Jacobian $J= k_0^{\phi_1}p_0^2x/(2 k_z^{\phi_1} k_z^{\phi_2}k_{\perp})$, and making use of the remaining Dirac delta, we arrive to a simplified expression for the decay probability,
\begin{equation}
    \mathcal{P}=\frac{1}{64\pi^2}\int \mathrm{d}x\mathrm{d}k_{\perp}^2\frac{|\mathcal{M}|^2}{k_z^{\phi_1}k_z^{\phi_2}}\,,
    \label{probabilidad reducida}
\end{equation} 
to be supplemented with the specific matrix element $\cal M$ for the interaction under consideration. Furthermore, in regard to the computation of the DM density via Eq.\,(\ref{expresion general densidad}), since the probability only depends on the $z$-component of the momenta of the incoming particle, we have
 \begin{equation}
      n_{\phi}=\frac{2}{\gamma_w v_w}\int_{-\infty}^{\infty} \frac{\textit{d}p_z}{(2\pi)^2} \, \mathcal{P}(p_z)\frac{\frac{\gamma_w p_z}{T_n}+1}{(\frac{\gamma_w}{T_n})^2} \, e^{\frac{\gamma_w}{T_n}v_wp_z}e^{-\frac{\gamma_w}{T_n}p_z}\,.
      \label{densidad reducida}
 \end{equation}
The detailed derivation of this expression can be found in Appendix \ref{A1}. Nonetheless, in general a numerical approach has been employed to compute the decay probability and the corresponding DM density via Eqs.\,(\ref{probabilidad reducida}) and (\ref{densidad reducida}) respectively. Subsequently, the DM abundance is obtained using Eq.~(\ref{parámetro densidad}).
\newline
By performing a numerical integration while fixing specific parameters $M,v,T_n$,  the DM abundance can be expressed as a function of $\gamma_w$.  It is worth noting  that the other parameters ($\Lambda,\lambda, T_f$)  can be adjusted post-integration since they only appear as a prefactor multiplying the integral. In this way, we generally find
\begin{equation}
    ^n\Omega_{BE} h^2=h^2\frac{\lambda^2g_{\star s0}}{g_{\star s}(T_{\text{f}})}\left(\frac{v}{\Lambda}\right)^{2(n-2)} \left(\frac{T_0}{T_{\text{f}}}\right)^3I_{BE}(\tilde{\gamma}_w,v,T_n,M),
    \label{densidad numerica}
\end{equation}
where the superscript $n$ stands for value of this parameter in Eq.\,(\ref{Lint}) and
\begin{equation}
\begin{split}
     I_{BE}=&\frac{ M}{\rho_c}\frac{T_n^2}{128\pi^4\gamma_w^3v_w}\int_{-\infty}^{\infty}\mathrm{d}p_z\left(1+\frac{p_z\gamma_w}{T_n}\right)\\
     \times & e^{-p_z(1-v_w)\frac{\gamma_w}{T_n}}\int_0^1\mathrm{d}x\int_0^{\infty}\mathrm{d}k_{\perp}\frac{\big|\mathcal{M}_{\star}\big|^2}{k_z^{\phi_1}k_z^{\phi_2}}
\end{split}
\end{equation}
with $\big|\mathcal{M}_{\star}\big|^2$ the rescaled squared matrix element resulting from removing the prefactor $\lambda^2(v/\Lambda)^{2(n-2)}$, i.e. $\big|\mathcal{M}\big|^2\equiv\lambda^2(v/\Lambda)^{2(n-2)}\big|\mathcal{M}_{\star}\big|^2$. It will be seen that the results are independent of the considered wall parametrization when $\gamma_w\gg M^2/v T_n$ as stated for other ansatzes in \cite{Azatov:2020ufh}. Additionally, we will see that in this regime is possible to obtain an analytic solution. Consequently, we define the {\it Lorentz reduced factor of the bubble wall} as  $\tilde{\gamma}_w=\gamma_w T_n v/M^2$ in order to distinguish both regimes. Namely, if $\tilde{\gamma}_w \gg 1$, one could have an analytic expression for the abundance, whereas for $\tilde{\gamma}_w<1$, a numerical approach is needed. Note that, since $\gamma_w\geq1$ then $\tilde{\gamma}_w\geq v T_n/M^2$.

 \section{DM PRODUCTION}\label{sec:BE MECHANISM}

Having outlined the basics of the BE mechanism, we proceed now to numerically investigate the renormalizable case for arbitrary wall velocities. Additionally, we revisit the analytical solution for the ultra-relativistic wall expansion originally considered in Ref.~\cite{Azatov:2021ifm}, where this scenario was first introduced.

\subsection{Four-dimensional interaction}\label{n=2}

For $n=2$, the interaction term shown in Eq.\,\eqref{Lint} reduces to the renormalizable expression $\mathcal{L}_{\text{int}}=\lambda \langle\Phi\rangle h\phi^2$. Assuming a weak-coupling constant $\lambda$, the corresponding matrix element $\mathcal{M}$ can be computed via Eq.~(\ref{matriz M}), taking (\ref{vertice}) as a prescription of $V(z)$. We obtain 
\begin{equation}
    \mathcal{M}=\lambda \left(\frac{v}{\Delta p_z}\right)\frac{e^{2i\sigma}-1}{2\sigma}\,,  \quad\quad\sigma=\frac{L_w\Delta p_z}{2}\,,
    \label{matriz M n2}
\end{equation}
and therefore
\begin{equation}
\big|\mathcal{M}\big|^2=\lambda ^2\left(\frac{v}{\Delta p_z}\right)^2 \textrm{sinc}^2\sigma \,.
\label{M cuadrado n2}
\end{equation}
Note that the obtained $\textrm{sinc}^2\sigma$ suppression factor in this expression is associated with the choice of a linear wall parametrization \cite{Azatov:2021ifm,Gouttenoire:2021kjv}. However, as argued in Ref.~\cite{Azatov:2021ifm},  the overall $n=2$ outcome in the relativistic regime $\Delta p_z L_w<1$ (or equivalently $p_0>M^2L_w$ or $\gamma_w>M^2 /vT_n$) is expected to remain consistent across other ansatzes, provided of course that their wall profiles exhibit similar behaviors to the linear one in the parameter regime under consideration.

\begin{figure}[]
\begin{centering}
\includegraphics[scale=0.42]{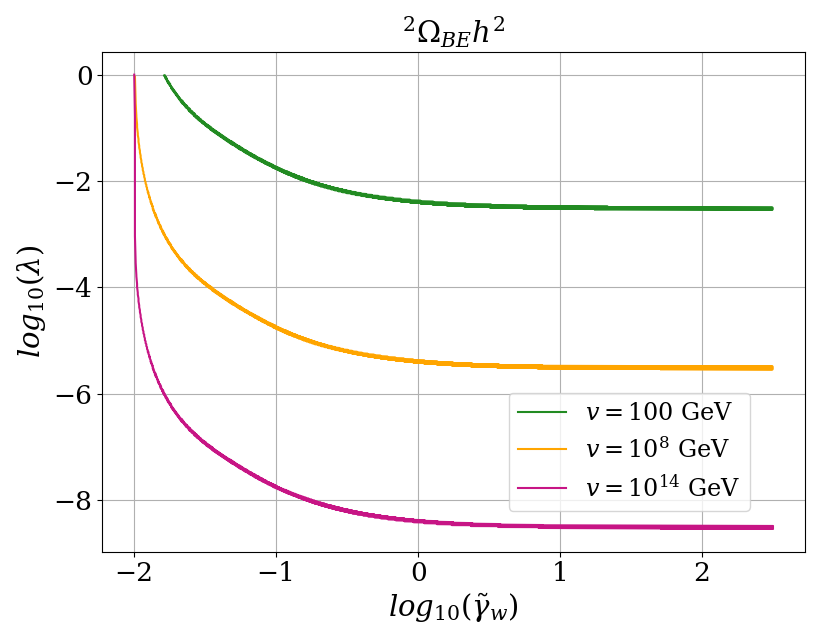}
\par\end{centering}
\caption{DM abundance in agreement with the observational value $\Omega_{\text{DM}}h^2=0.1200\pm0.0012$ \cite{Planck:2018vyg} taking a renormalizable scalar-scalar interaction ($n=2$) in terms of the coupling $\lambda$ and the
{\it Lorentz reduced factor of the bubble wall} $\tilde{\gamma}_w$. The presented numerical results are for PTs at different scales, taking the nucleation temperature and the final temperature equal to the vev inside the bubble ($T_n=T_f=v$) and a mass for the DM particle one order of magnitude heavier ($M=10v$).}
\label{Figura1}
\end{figure}

Making use of this squared matrix element we numerically derive the DM abundance following the procedure described in the previous section, arriving to an expression like the one in Eq.\,(\ref{densidad numerica}). Thus, in Figure \ref{Figura1}, we present the parameter space allowing for DM production in agreement with the observational value $\Omega_{\text{DM}}h^2=0.1200\pm0.0012$ \cite{Planck:2018vyg} for PTs occurring at different scales. Notably, in the previously unexplored low-velocity regime, it is possible to achieve DM production consistent with observations, thereby expanding the range of possibilities offered by this mechanism. The value $\tilde{\gamma}_w\simeq1$ marks the boundary between two distinct regimes: For $\tilde{\gamma}_w\lesssim1$ the DM abundance depends on the Lorentz factor. This is reflected in Fig.\,\ref{Figura1} through the need of different values of the coupling constant ($\lambda$) in this regime. For $\tilde{\gamma}_w\gtrsim1$, the abundance becomes independent of velocity, and $\lambda$ adopts a fixed value. Furthermore, for PTs at scales higher than the EW scale, it is possible to encounter scenarios with $T_f>T_n$ and suppressed the DM abundance. However, given the range of $\lambda$ values available in these scenarios, this suppression can be always compensated by selecting higher values of the coupling constant.

 Having studied numerically the DM production, 
 considering all possible regimes for the Lorentz factor, we will focus now on the ultra-relativistic regime ($\tilde{\gamma}_w>1$), where it is possible to derive an analytic expression for the DM abundance, as already studied for this renormalizable case in \cite{Azatov:2020ufh}. Having derived the matrix element in Eq.\,(\ref{matriz M n2}) it is possible to derive the probability of decay via Eq.\,(\ref{probabilidad reducida}) imposing that $\Delta p_z L_w<1$ ($\tilde{\gamma}_w>1$), $p_0^2 (1-x)^2\gg k_{\perp}^2+M^2$ and $p_0^2x^2 \gg k_{\perp}^2+M^2$ (see Appendix \ref{A0} for the details of the computation). Thus, we have
\begin{equation}
    \mathcal{P}=\frac{\lambda^2}{96\pi^2}\left(\frac{ v}{M}\right)^2\Theta(p_0-2M)\Theta(p_0-M^2L_w)\,.
    \label{probabilidad n=2}
\end{equation}
Note that this equation corrects the result of \cite{Azatov:2021ifm} by a factor $1/4$, with the latter Heaviside function ensuring the regime $\Delta p_z L_w<1$.  Moreover, since $L_w\sim1/v$ \cite{Azatov:2021ifm}, the assumed hierarchy $M\gg v$ implies $M^2L_w>2M$. Given the probability we can derive the DM density via Eq.\,(\ref{probabilidad reducida}) and with it the corresponding relic abundance through Eq.\,(\ref{parámetro densidad}). Namely,
\begin{equation}
    n_{\phi}\simeq \frac{\lambda^2  T_{\text{n}}^3}{48\pi^4}\left(\frac{v}{M} \right)^2\,,
\label{densidad n2 aprox}
\end{equation}
and
\begin{equation}
     ^2\Omega_{\text{BE}} h^2\simeq\frac{200\Upsilon}{48\pi^4} \frac{\lambda ^2}{g_{\star s}} \frac{v}{M}\frac{v}{200\,\text{GeV}}\left(\frac{T_{\text{n}}}{T_{\text{f}}}\right)^3\equiv^2\overline{\Omega}_{\text{BE}}h^2\,,
     \label{parametro densidad n=2}
\end{equation}
that is in concordance with the numerical results in the large $\tilde{\gamma}_w$ regime.

\begin{figure}[]
\begin{centering}
\includegraphics[scale=0.42]{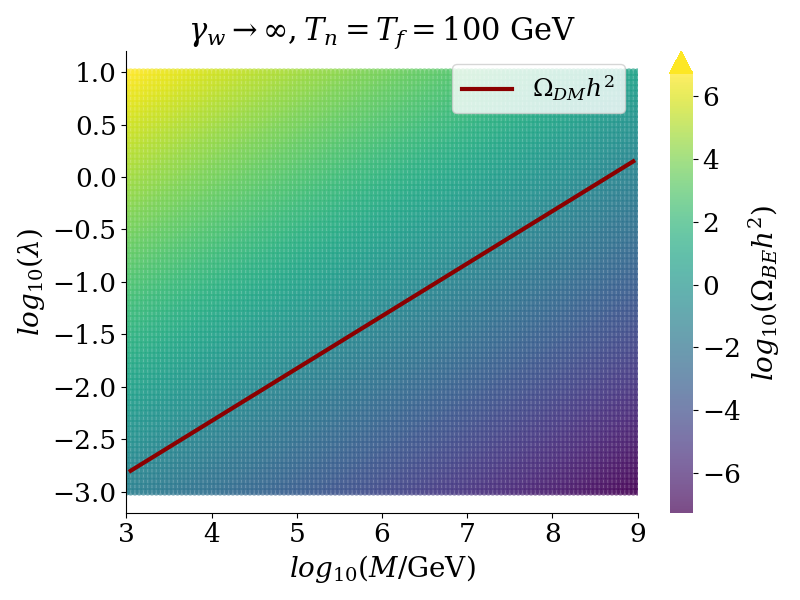}
\par\end{centering}
\begin{centering}
\includegraphics[scale=0.42]{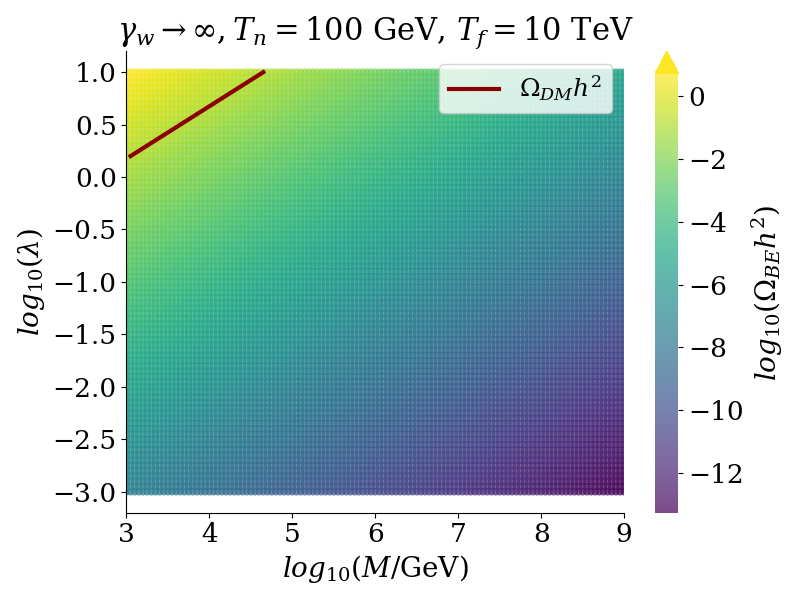}
\par\end{centering}
\caption{DM relic abundance for a renormalizable scalar-scalar interaction ($n=2$) in the ultra-relativistic limit in terms of the 
DM mass $M$ and the coupling $\lambda$. The vev inside the bubble is fixed to $v=200$ GeV and the nucleation temperature to $T_{\text{n}}=100$ GeV, whereas the final temperature is taken as $T_{\text{f}}=100$ GeV in the upper panel and $T_{\text{f}}=10$ TeV for the lower panel. The red line stands for the points in agreement with the observational value, $\Omega_{\text{DM}}h^2=0.1200\pm0.0012$ \cite{Planck:2018vyg}. Note that in the lower panel there is no parameter fulfilling the weak-coupling regime ($\lambda \lesssim 1$).}
\label{Figura2}
\end{figure}

As shown in the upper panel of Figure~\ref{Figura2}, choosing parameter values at the electroweak (EW) scale allow for DM production consistent with the observational value  $\Omega_{\text{DM}}h^2=0.1200\pm0.0012$ \cite{Planck:2018vyg} for a wide range of couplings $\lambda$. This, in turn, permits high DM candidate masses that are typically forbidden in other DM production mechanisms \cite{Griest:1989wd}. Note, however, that this renormalizable scenario exhibits a strong dependence on the final temperature $T_f$, as was already observed for transitions at this scale at low velocities. Specifically, as $T_f$ increases, the DM abundance decreases, reducing the parameter space that aligns with observational constraints, as evident in the lower panel of Figure \ref{Figura2}. Moreover, for $T_{\text{f}}>100\,T_n$,  there are no values within the considered ranges of $M$ and $\lambda$ that yield sufficient DM production.

The complete numerical results for PTs at the previously discussed scales will be studied below together with the results for the non-renormalizable cases. These results will explicitly demonstrate that the DM abundance is independent of the wall velocity in the ultra-relativistic regime, in concordance with the analytical approximation. 

\subsection{Non-renormalizable interactions}\label{subsection:B}

\begin{figure}[]
\begin{centering}
\includegraphics[scale=0.42]{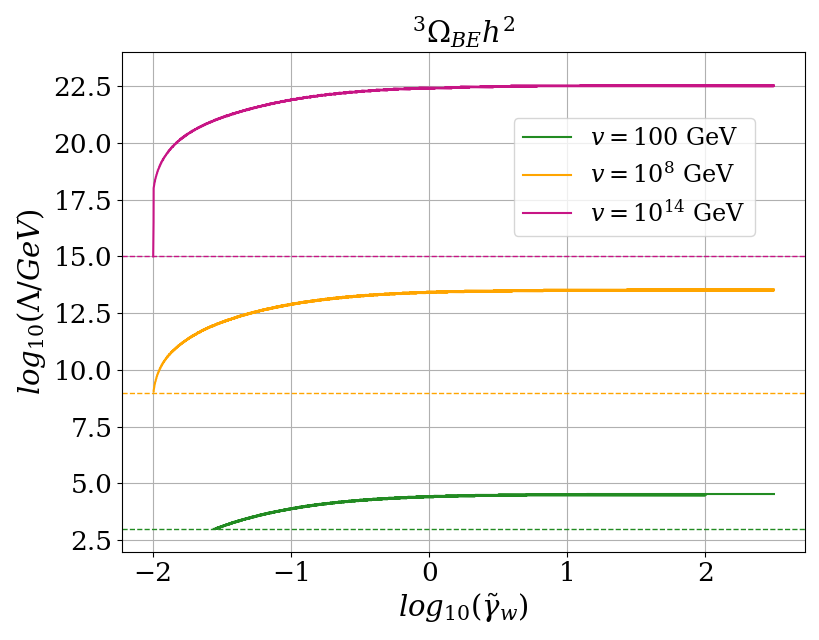}
\par\end{centering}
\begin{centering}
\includegraphics[scale=0.42]{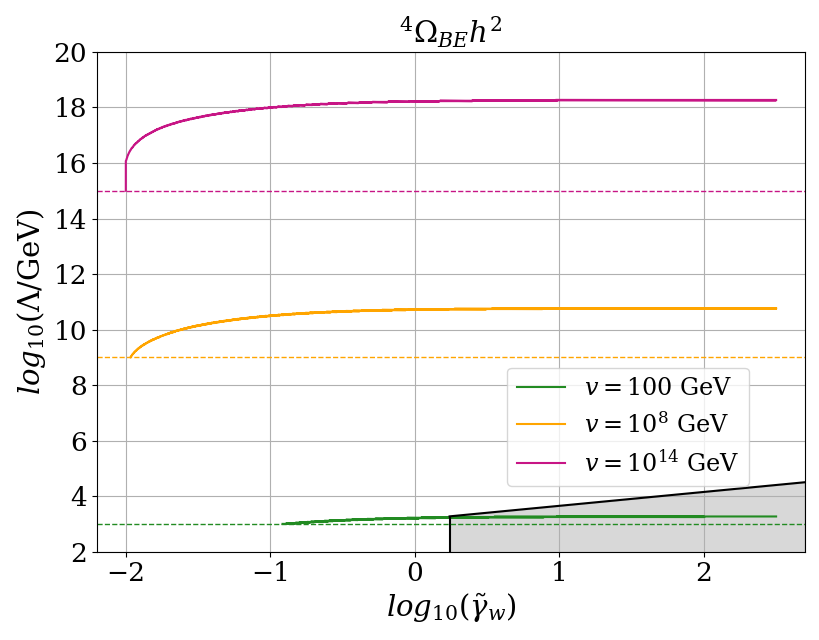}
\par\end{centering}
\caption{Same as Figure\,\ref{Figura1} but for $n=3$ (upper panel) and $n=4$ (lower panel). For this reason, the results are shown in terms of the cutoff scale $\Lambda$. The shaded region corresponds to the parameter space where the condition $2p_0v<\Lambda^2$ is violated for $v=100$ GeV, rendering it excluded due to the breakdown of the effective field theory. In the other cases, the presented lines in agreement with the observational value fulfill this condition.}
\label{Figura3}
\end{figure}

The previous scenario represents the simplest realization of the BE mechanism, enabling DM production over a wide range of Lorentz factor values associated with bubble expansion, as well as the generation of heavy DM candidates in the relativistic regime. However, it is worth exploring more general non-renormalizable interactions, which could expand the parameter space and introduce distinct dependencies on the model parameters. To this end, we will analyze the BE mechanism with an interaction as described in Eq.~(\ref{Lint}), now considering the cases $n=3$ and $n=4$.  For clarity, we will denote expressions specific to the cases $n=3$ or $n=4$ with a superscript $3$ or $4$, respectively. Specifically, we define $^3\mathcal{L}_{\mathrm{int}}=\langle\Phi\rangle^2 h\phi^2/\Lambda$ and $^4\mathcal{L}_{\text{int}}=\langle\Phi\rangle^3 h\phi^2/\Lambda^2$, where the coupling constant $\lambda$ has been conventionally absorbed into the definition of the cutoff scale. For consistency, this cutoff scale is assumed to be the largest energy scale in the theory.\footnote{Note that for non-renormalizable cases, additional interactions arising from Eq.\,\eqref{Lint} may eventually become significant in the $\tilde{\gamma}_w\gg1$ regime, potentially affecting the range of validity of some presented figures. Nevertheless, we report the results in this regime while considering only the specified interactions, enabling a direct comparison with the renormalizable case and the existing literature on the subject. Importantly, our results remain fully applicable for any $\gamma_w$, as long as the insertions are interpreted as a modification of the profile in the renormalizable case.}
\newline
To compute the DM abundance, we carry out a procedure analogous to the previous case, calculating the matrix element $\mathcal{M}$ via Eq.~(\ref{matriz M}), with $V(z)$ given again by (\ref{vertice}). We obtain 
\begin{equation}
    ^3\mathcal{M}=2i\lambda\left(\frac{v}{\Delta p_z}\right)^2\frac{1}{L_w\Lambda}\left[\frac{e^{2i\sigma}-1}{2\sigma}-ie^{2i\sigma} \right]\,,
    \label{M n3}
\end{equation}
and 

\begin{equation}
\begin{split}
    \hspace{-3mm}
^3\big|\mathcal{M} \big|^2=4\lambda^2 \left( \frac{v}{\Delta p_z} \right)^4 \left(\frac{1}{\Lambda L_w}\right)^2\left(\text{sinc}^2\sigma+1-2\,  \text{sinc\,}2\sigma\,\right),
\end{split}
\label{Matriz M n3}
\end{equation}
for the five-dimensional interaction, and 
\begin{equation}
\begin{split}
    ^4\mathcal{M}=6\lambda \left(\frac{v}{\Delta p_z}\right)^3\left(\frac{1}{\Lambda L_w}\right)^2 \left[\frac{1-e^{2i\sigma}}{2\sigma}+e^{2i\sigma}\left(\sigma+i\right)\right]\,,
\end{split}
\end{equation}
or equivalently
\begin{equation}  \hspace{-3mm}  
\begin{split}
 ^4\big|\mathcal{M}\big|^2=36\lambda^2 \left(\frac{v}{ \Delta p_z}\right)^6\left(\frac{1}{\Lambda  L_w}\right)^4 \times \\ \left[ \text{sinc}^2\sigma+1-2 \,\text{sinc}\, 2 \sigma+\sigma ^2(1-2
   \text{sinc}^2\sigma)\right]\,,
\end{split}
\label{Matriz Mn4}
\end{equation}
for the six-dimensional one. Using these squared matrix elements, we numerically compute the DM abundance by following the same procedure as in the previous section. Specifically, we use Eq.\,(\ref{probabilidad reducida}) to calculate the probability and Eq.\,(\ref{densidad reducida}) to determine the DM density. This leads to an analogous expression to that presented in Eq.\,(\ref{densidad numerica}). 
\newline
The parameter space consistent with the observed DM abundance for PTs ocurring at various energy scales is illustrated in Figure~\ref{Figura3}. Once again, we find that the correct DM abundance can be achieved regardless of the value of  $\tilde{\gamma}_w$,  including in the previously unexplored $\tilde{\gamma}_w<1$ regime. As in the renormalizable case, the abundance becomes independent of the wall velocity in the the $\tilde{\gamma}_w \gg 1$ regime. As before, for PTs at scales higher than the EW scale, the suppression effect caused by a final temperature higher than the nucleation temperature ($T_f>T_n$) can be offset by adjusting the associated values of the cutoff scale. Note, however, that, in order to maintain the validity of the effective field theory one must require $\Lambda \gtrsim M$, being this the reason why achieving as low velocities at the the EW scale as those accessible at other scales is not generically possible. For the same purpose, we require that the center-of-mass energy of the particle-bubble system remains below the cutoff scale ($s\simeq2p_0v<\Lambda^2$). This condition excludes a portion of the parameter space for a PT at the EW scale in the 6-dimensional case.

Having numerically studied the DM production, we now derive an analytical expression for the DM abundance in the ultra-relativistic regime $\tilde{\gamma}_w \gg 1$.  The details of the computation are provided in Appendix~\ref{A1} for the 5-dimensional case and Appendix~\ref{A2} for the 6-dimensional case. From the already presented matrix element in Eqs.\,(\ref{Matriz M n3},\ref{Matriz Mn4}) we calculate the probability of decay via Eq.\,(\ref{probabilidad reducida}). By making the same assumptions as in the previous section, while additionally ensuring that the center-of-mass energy remains below the cutoff scale ($s\simeq2p_0v<\Lambda^2$)\footnote{This can be deduced from the fact that the variation of momentum due to the bubble is limited by the scale of the vev for an efficient production of the DM particles: $\Delta p_z\lesssim v$ (see Eq.\,(\ref{M cuadrado n2}), Eq.\,(\ref{Matriz M n3}) and Eq.\,(\ref{Matriz Mn4})).} to preserve the validity of the effective field theory, we obtain
\begin{equation}
\begin{split}
     ^3\mathcal{P}=\frac{\lambda^2}{24 \pi^2}  \left(\frac{v}{M}\right)^4 \left( \frac{1}{\Lambda L_w}\right)^2
     \Theta(a)\Theta(b) \bigg[\frac{2}{35} \left(\frac{p_0}{M}\right)^2\times\\\big(1+\Theta(c)-2\Theta(d)\big)
        +\frac{\left(M L_w\right)^2}{3}\bigg(\Theta(d)-\frac{1}{4}\Theta(c)\bigg)\bigg]\,,
\end{split}  
    \label{probabilidad n=3}
\end{equation}
and 
\begin{equation}\label{Pn4}
    \begin{split}
        ^4P=\frac{3\lambda^2}{8\pi^2}\left(\frac{v}{M}\right)^6 \left(\frac{1}{\Lambda  L_w}\right)^4 \Theta(a)\Theta(b) \bigg[\frac{8}{1155}\left(\frac{p_0}{M}\right)^4 \\\left(1+\Theta(c)-2\Theta(d)\right)+\frac{(p_0 L_w)^2}{70} \bigg(1-\frac{7}{3}\Theta(c)\\ +\frac{4}{3}\Theta(d)\bigg)
        +\frac{1}{12}(L_wM)^4\left(\frac{25}{48}\Theta(c)-\frac{1}{5}\Theta(d)\right)
    \bigg]\,,
    \end{split}
\end{equation}
 with $a=(p_0-2M)$, $b=(\Lambda^2-2p_0v)$, $c=(p_0-M^2L_w)$ and $d=(p_0-2M^2L_w)$. In the computation, the $\text{sinc}\,(\sigma)$ functions appearing in the matrix elements have been replaced with their Taylor expansions. To ensure the validity of this expansion, appropiate Heaviside functions were introduced, enforcing the condition $\Delta p_z L_w<1$ ($\tilde{\gamma}_w \gg 1$). Note that in the considered heavy DM limit $M\gg v$ \cite{Azatov:2021ifm},  the inequalities $2M^2L_w>M^2L_w>2M$  hold, effectively relating the arguments of the Heaviside functions.

Given the above probabilities we can derive the DM density via Eq.~(\ref{expresion general densidad}) and with it the corresponding relic abundance through Eq.~(\ref{parámetro densidad}). We present the complete expressions in Table \ref{table:expresiones completas}, only quoting here the leading-order terms in $\tilde{\gamma}_w\gg1$, namely
\begin{equation}
^3\Omega_{\text{BE}}h^2  \simeq \frac{v^2}{\Lambda^2}\left(1-\frac{1}{\tilde{\gamma}_w^2}\left(\frac{51}{280}+\frac{8v^4}{35M^4}\right)\right) {}^2\overline{\Omega}_{\text{BE}} h^2\,,
\label{parametro densidadn3}
\end{equation}
for the 5-dimensional case ($n=3$) and 
\begin{equation}
\begin{split}
    ^4\Omega_{\text{BE}}h^2\simeq &\frac{ v^{4}}{\Lambda^4}\bigg[1+\frac{1}{\tilde{\gamma}_w^2}\bigg(\frac{5623}{7392}\\&-\frac{18v^4}{35M^4}-\frac{256v^6}{385M^6}\bigg)\bigg]{}^2\overline{\Omega}_{\text{BE}} h^2 \,,
\end{split}
    \label{parametro densidadn4}
\end{equation}
for the 6-dimensional one ($n=4$). Note also  that the relationship between $^3n_{\phi}$ and $^4n_{\phi}$ with $^2n_{\phi}$ is the same as the presented for their respective abundances.

The leading order analytical result obtained for $\tilde{\gamma}_w \gg 1$  is entirely analogous to this derived for the renormalizable case, apart from the rescaling factors $v^2/\Lambda^2$, $v^4/\Lambda^4$. Recalling that, aside from these factors, the non-renormalizable interaction under consideration can be interpreted as a reparametrization of the wall, this indicates that the results are independent of the wall shape in this limit for the ansatzes considered. This conclusion aligns with previous findings for other parameterizations studied in Ref.~\cite{Azatov:2020ufh}. From the chosen reference frame, if an incoming particle approaches the wall at high velocity, its decay is unaffected by the wall's shape, as the particle is unable to resolve its details. In this ultra-relativistic regime, the different abundances provided by distinct bubble wall shapes are suppressed by the square inverse value of the {\it Lorentz reduced factor of the bubble wall} $\tilde{\gamma}_w$. The same suppression affects the differences in the abundances among distinct interaction temrs, beyond the commented rescaling factors provided by the vacuum insertions of the scalar field.

\begin{figure}[]
\begin{centering}
\includegraphics[scale=0.42]{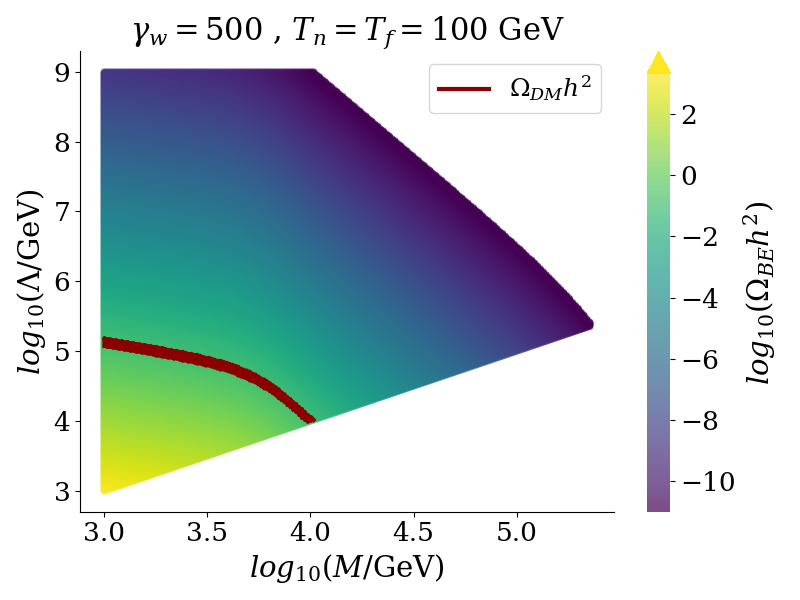}
\par\end{centering}
\begin{centering}
\includegraphics[scale=0.42]{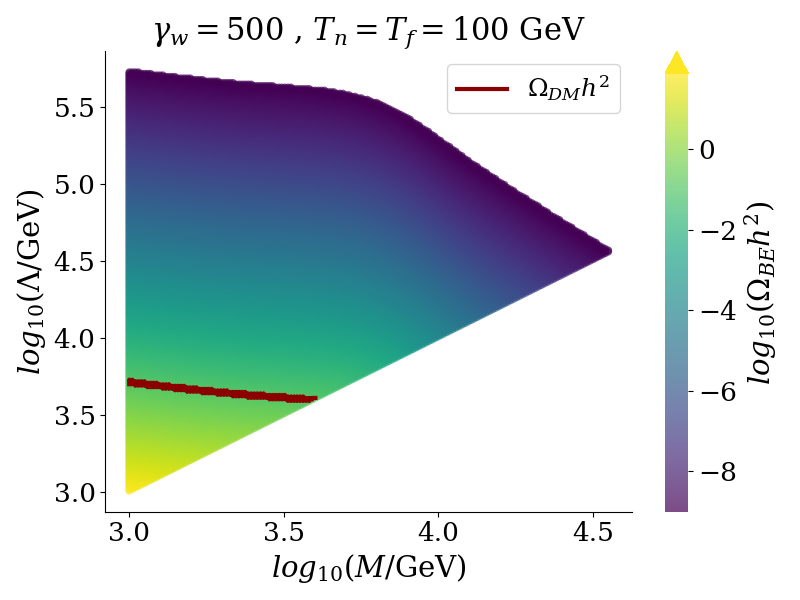} 
\par\end{centering}
\caption{Same as for the upper panel in Figure \ref{Figura2} but for a non-renormalizable interaction with $n=3$ (upper panel) and $n=4$ (lower panel) with $\gamma_w=500$. Therefore, it shows the abundance in terms of the cutoff scale $\Lambda$.}
\label{Figura4}
\end{figure}

\begin{figure}[h]
\begin{centering}
\includegraphics[scale=0.39]{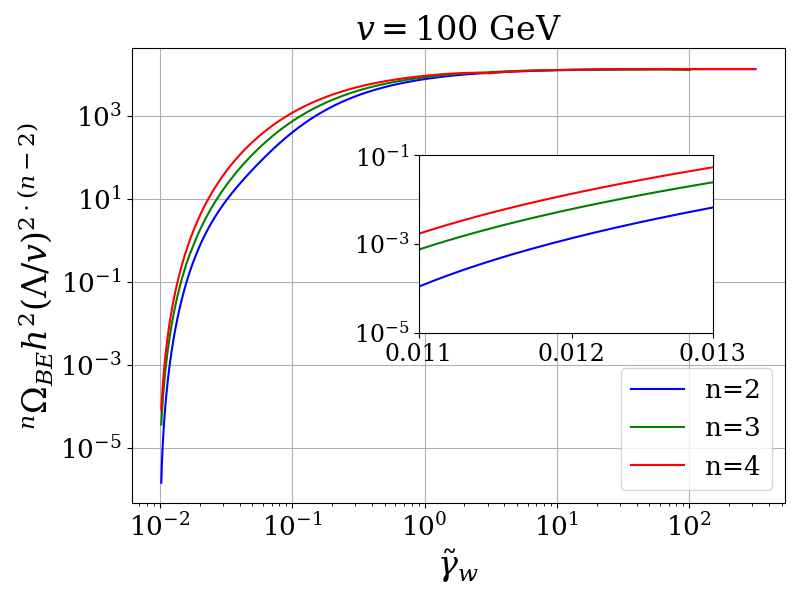}
\par\end{centering}
\begin{centering}
\includegraphics[scale=0.39]{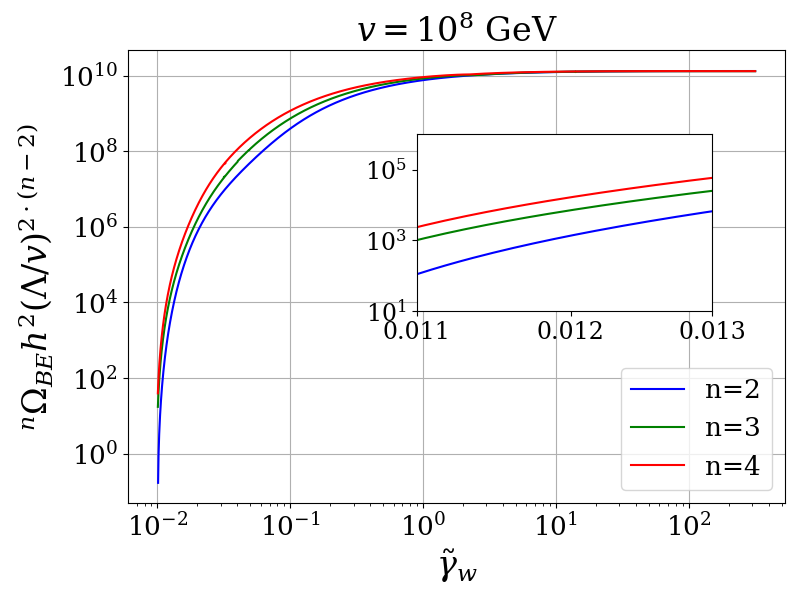}
\par\end{centering}
\begin{centering}
\includegraphics[scale=0.39]{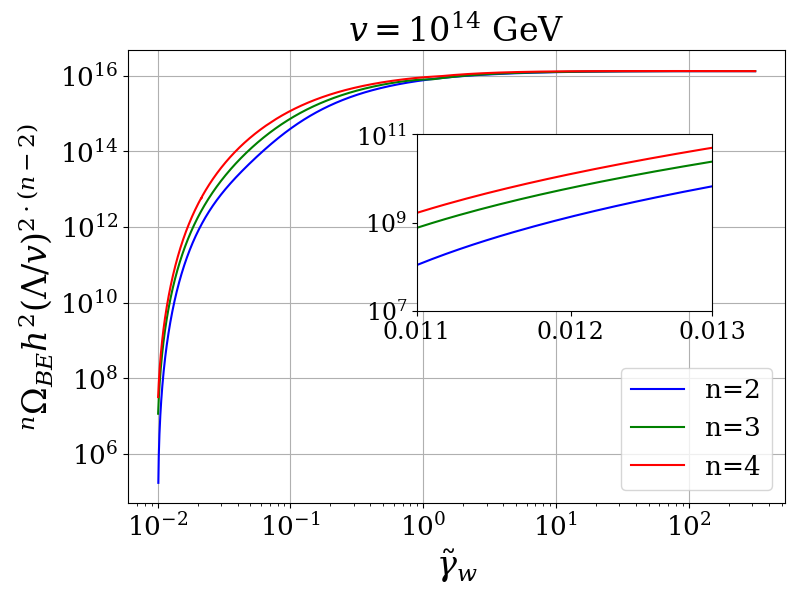}
\par\end{centering}
\caption{Comparison of the numerical results for the DM abundance produced in a PT at different scales in terms of $\tilde{\gamma}_w$ for the different operators considered. Non-renormalizable cases are scaled with the appropriate vacuum insertions. For the presented numerical results, we fix $T_n=T_f=v$ and $M=10v$.}
\label{Figura 5}
\end{figure}

The parameter space allowed by $^3\Omega_{\mathrm{BE}}h^2$ ($^4\Omega_{\mathrm{BE}}h^2$) for a FOPT at the EW scale is displayed in the upper (lower) panel of Figure \ref{Figura4}. When studying the values in agreement with the observations, we use the complete expression for $^{3,4}\Omega_{\text{BE}}h^2$ presented in Table \ref{table:expresiones completas}.  The lower cut corresponds to the limiting mass value consistent with the effective field theory description, $M\simeq\Lambda$. It can be seen that the parameter space is reduced as compared to the renormalizable case due to the suppression factor $(v/\Lambda)^{2,4}$, which limits the achievable mass values. As a result, masses as heavy as in the renormalizable case are not possible under the conditions shown in Figure~\ref{Figura4}. In general, we find that, aside from the aforementioned rescaling factors, the DM abundance  in the limit $\tilde{\gamma}_w\gg 1$ takes always the same form, exhibiting complete independence from the parametrization of the wall when the expansion velocity is sufficiently high.

Finally, in Figure\,\ref{Figura 5} we present the complete numerical results alongside those for the renormalizable case for PTs at the scales previously considered. To facilitate a comparison of the different operators, we have omitted the rescaling factors in the non-renormalizable cases, introducing additional factors of $\Lambda^2/v^2$, $\Lambda^4/v^4$ when needed.  It is important to recall that, in the absence of these factors, the vacuum expectation value (vev) insertions can be interpreted as a reparametrization of the wall shape. Furthermore, Figure\,\ref{Figura 5} provides 
a comparison of the impact of the different operators considered (corresponding to different wall parametrizations), particularly in the $\tilde{\gamma}_w\ll1$ regime, where deriving an analytical approximation has not been possible.

In concordance with the analytical computation, the leading order results in the ultra-relativistic regime $\tilde{\gamma}_w \gg 1$, are independent of the wall velocity and are identical for all three operators, aside from the aforementioned rescaling factors.  Furthermore, in the $\tilde{\gamma}_w<1$, the abundance behavior for the different operators is qualitatively similar, with differences that are limited to an order of magnitude. This is in contrast to initial expectations, as one might have assumed that the wall shape could influence the results in a much more important way at low velocities of the incoming particle. Nonetheless, the dimension of the operators still play a role in the DM abundance. Additionally, we observe that the higher the scale at which the PT occurs, the greater the DM abundance, which is consistent with the predictions from the analytical approximation. Finally, we do not expect differences beyond rescaling factors when considering other profiles with non-renormalizable operators, as these can effectively be interpreted as linear profiles in higher-dimensional operators.

Thus, it has been seen how, although the ultra-relativistic regime provides facilities in the computation, allowing for analytical results and independent of the wall profile, there is no need to be restricted to this regime. On the contrary, we have found that the case $\tilde{\gamma}_w \lesssim 1$ is also very interesting since it provides an alternative and considerable parameter space for the DM abundance in agreement with the observational value in the low velocities regime. Furthermore, the different numbers of vev-insertions do not yield for important qualitative differences in the behavior of the abundance apart from the rescaling factors, even in the $\tilde{\gamma}_w<1$ regime. In any case, the exact results depend on the interaction term, what can be interpreted as an signal of the wall shape in the observable DM production wihtin this regime.
\section{GWs SIGNAL}\label{sec: GW SIGNAL}

One of the most interesting features of our DM production via expanding bubbles is the associated generation of sizable GWs, as this signal could eventually serve as indirect evidence for it.

GWs in FOPT have been a widely studied topic in recent years, being their current understanding essentially due to hydrodynamic and scalar field simulations \cite{GWCaprini:2019egz,Gould:GW,GWCutting:2018tjt,GWHindmarsh:2017gnf,Hindmarsh:GW}. In the scenarios under consideration, we can identify a priori three possible contributions to the gravitational wave spectrum: the \textit{scalar field} contribution due to collisions between bubbles, the \textit{sound waves} contribution due to sound waves generated by the energy transfer from $h$ to the plasma, and the \textit{turbulent} contribution due to nonlinear effects in the turbulent motion of the plasma. However, due to the large uncertainties it presents and since it is expected to be subdominant \cite{GWCaprini:2019egz}, we will not consider the turbulent contribution. To parametrize the relative weight of the two remaining contributions, we introduce the effective parameters, 
\begin{equation}
k_{\text{\rm wall}}=\frac{E_{\text{\rm wall}}}{E_{\text{total}}}\,, \hspace{10mm} k_{\text{fluid}}=1-k_{\text{\rm wall}}\,,
\end{equation}
governing respectively the energy distribution between the wall motion and the plasma excitation, with $E_{wall}$ the wall kinetic energy.

Regarding the velocity of bubbles, we can distinguish two regimes: the \textit{terminal velocity} regime, in which the friction is sufficient to slow down the bubble wall so it reaches a constant velocity, and the \textit{runaway} regime, in which the energy released is such that the wall continues to accelerate until collision, reaching ultra-relativistic velocities. In the former case, only sound waves contribute to the GW production since, once the terminal velocity is reached, the portion of the energy stored in the wall starts to decrease as the inverse of the bubble radius, so that $k_{\text{fluid}}=1$. On the other hand, in the latter case we have $k_{\text{\rm wall}}=1-\alpha_{\infty}/\alpha$, 
with $\alpha$ the transition strength parameter introduced in Section \ref{sec:walldynamics}, and $\alpha_{\infty}$ its value from which we are in the runaway regime \cite{AinftyEspinosa:2010hh}. Thus, both the \textit{scalar} and \textit{sound waves} contributions are a priori important in this case. 

The \textit{scalar field} or \textit{bubble collision} GW contribution following from numerical simulations takes the form \cite{GWCutting:2018tjt}
\begin{equation}
    \frac{\textit{d}\Omega_{\phi}h^2}{\textit{d}ln(f)}=4.7\cdot10^{-8}\left(\frac{100}{g_{*s}}\right)^{1/3}(H_{\text{f}}R_*)^2\left(\frac{k_{\text{\rm wall}}\alpha}{1+\alpha}\right)^2S_{\text{\rm wall}},
\end{equation}
with $g_{*s}$ evaluated at the time of GWs generation,  $H_{\text{f}}$ the Hubble parameter at the final temperature and 
\begin{equation}
    R_*=\frac{(8\pi)^{1/3}v_w}{\beta}\,,
\end{equation}
the size of the bubble at the collision. Here $S_{\text{\rm wall}}$ stands for the numerically-fitted spectral function
\begin{equation}
    S_{\text{\rm wall}}=\frac{(a+b)^cf_p^bf^a}{(bf_p^{\frac{a+b}{c}}+af^{\frac{a+b}{c}})^c}\,,
\end{equation}
with $a=3$, $b=1.51$, $c=2.18$, $f$ the frequency and
\begin{equation}
    f_p=5.28\cdot10^{-4}\frac{T_{\text{f}}}{100 \, \text{GeV}}\left(\frac{g_*}{100}\right)^{1/6}\frac{1}{2\pi R_*H_{\text{f}}}\,,
\end{equation}
the peak frequency.

On the other hand, the \textit{sound waves} GW contribution in the prescription of Ref.~\cite{GWHindmarsh:2017gnf} reads
\begin{equation}
    \frac{\textit{d}\Omega_{sw}h^2}{\textit{d}ln(f)}=\left\lbrace\begin{array}{cr}
        0.678h^2F_{\text{\text{sw}}}G\Tilde{\Omega}_{\text{sw}}C(s), & \text{if}\quad \frac{H_{\text{f}}R_*}{K^{1/2}}>1\,;  \\
        0.678h^2F_{\text{sw}}\Bar{G}\Tilde{\Omega}_{\text{sw}}C(s), & \text{if}\quad \frac{H_{\text{f}}R_*}{K^{1/2}}<1\,. 
    \end{array}\right. \,
    \label{espectro ondas sonido}
\end{equation}
Here, we have $F_{\text{sw}}=3.57\cdot10^{-5}(100/g_*)^{1/3}$, $G=K^2H_{\text{f}}R_*/c_s$ and $\Bar{G}=K^{3/2}(H_{\text{f}}R_*/c_s)^2$, with $c_s$ the speed of sound and
\begin{equation}
    K\approx \frac{3}{4}\frac{k_{\text{sw}}\alpha}{1+\alpha}, \quad k_{\text{sw}}=k_{\text{fluid}}\frac{\alpha}{0.73+0.083\sqrt{\alpha}+\alpha}\,,
\end{equation}
\begin{figure}[]
\begin{centering}
\includegraphics[scale=0.32]{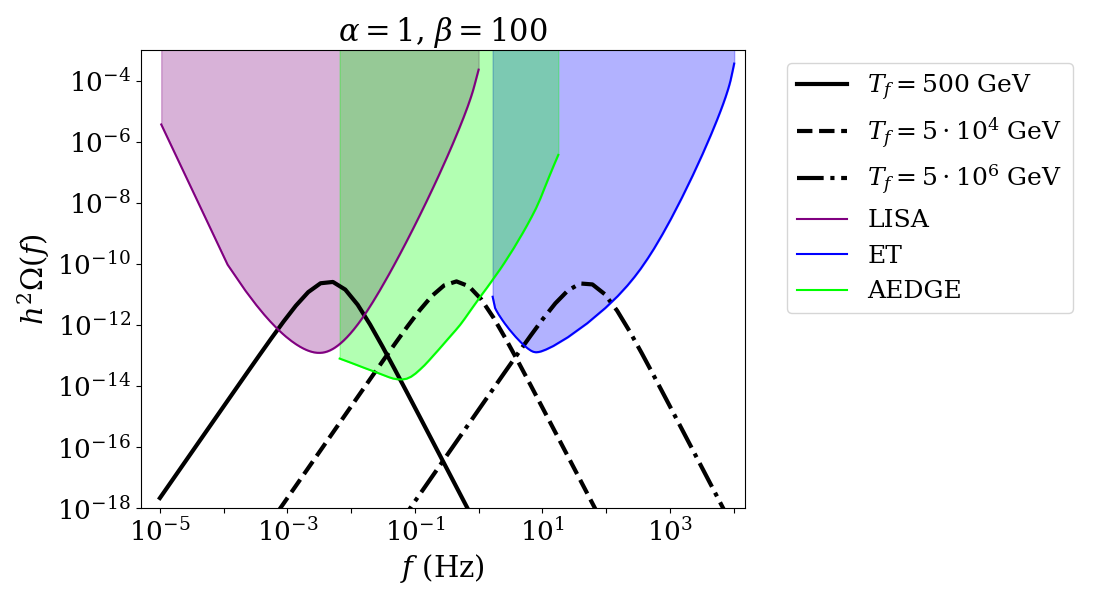} 
\par\end{centering}
\caption{GW spectra for different values of the
final temperature $T_{\text{f}}$ and fixed values of the
transition strength parameter $\alpha$ and the transition rate parameter $\beta$ in the terminal velocity regime ($v_w=0.95$). We are always assuming $\alpha_{\infty}=0.001$. The presented value of $\beta$ is in units of the Hubble parameter at the final temperature $H_{\text{f}}$.  LISA \cite{LISA}, ET \cite{ET} and AEDGE \cite{AEDGE:2019nxb} sensitivity curves are also presented.}
\label{figura espectros}
\end{figure}
\noindent where, for the case of runaway regime, we have to substitute $\alpha$ with $\alpha_{\infty}$. The spectral shape $C(s)$ is numerically found to be
\begin{equation}
    C(s)=s^3\left(\frac{7}{4+3s^2}\right)^{7/2}\,, \quad  \quad s\equiv \frac{f}{f_p}\,,
\end{equation}
with peak frequency
\begin{equation}\label{eq:fp}
    f_p\simeq 26\cdot10^{-6}\frac{1}{H_{\text{f}}R_*}\frac{z_p}{10}\frac{T_{\text{f}}}{100\,\text{GeV}}\left(\frac{g_*}{100}\right)^{1/6} \text{Hz}\,,
\end{equation}
and peak angular frequency $z_p=(kR_*)_{\text{max}}$ generally around 10. Last, numerical simulations give $\Tilde{\Omega}_{\text{sw}}\approx 10^{-2}$ \cite{GWHindmarsh:2017gnf}.

The GW spectra resulting from the above expressions is shown in Figure \ref{figura espectros} for $\alpha_{\infty}=0.001$, typical values of $\alpha$ and $\beta$, and different values of the final FOPT temperature $T_f$, which as given by Eq.~\eqref{eq:fp} depends linearly on the peak frequency $f_p$. As shown in this plot, a FOPT with final temperature above the EW scale ($100\,\text{GeV}$) and below $T_f=10^7\, \text{GeV}$ is well within the reach of future GWs experiments. Note also that, although we have concentrated in a \textit{terminal velocity regime}, the spectra for \textit{runaway regime} would be also very similar. 

\section{CONCLUSIONS AND OUTLOOK}\label{sec:CONCLUSIONS}

Cosmological first-order phase transitions provide a natural way of producing massive dark matter particles in the very early Universe.  In this paper, we have considered the \textit{Bubble Expansion} mechanism for the production of scalar DM candidates, focusing on the non-relativistic regime of expansion and exploring the role of non-renormalizable interactions between the dark sector and the field undergoing the transition, while remaining otherwise agnostic about the specific effective potential responsible for the process. Our results complement previous studies in the literature dealing with renormalizable scalar interactions in the ultra-relativistic regime \cite{Baldes:2022oev}, Proca-like couplings \cite{Ai:2024ikj} and non-renormalizable operators mediating the production of heavy fermions and gauge bosons \cite{Azatov:2024crd}.

In contrast to previous works, we have explored the DM production in the non-relativistic regime, explicitly demonstrating that a DM abundance consistent with observations is achievable even at small wall velocities, thereby expanding the parameter space of the original scenario. Additionally, we have examined the effect of non-renormalizable interactions, considering different numbers of vev insertions. It was found that, apart from rescaling factors arising from the various operator dimensions in Eq.\,(\ref{Lint}), there are no significant differences in the qualitatively produced abundance. Indeed, the leading order results are the same in the ultra-relativistic regime $\tilde{\gamma}_w \gg 1$, where we have found an analytical expression for the DM abundance that is independent of the wall shape. We have also computed the first correction in the Lorentz reduced factor of the bubble wall for the non-renormalizable interactions. They are proportional to ${\tilde{\gamma}_w}^{-2}$ and are sensitive to the wall profile. In any case, the dependence in the wall shape is maximize for $\tilde{\gamma}_w<1$, where the non-renormalizable interactions show differences upto one order of magnitude among them (in addition to the vev insertion terms).
Interestingly enough, for final temperatures in the range $100\, {\rm GeV}\lesssim T_f\lesssim 10^7\, \text{GeV}$, the GW spectrum resulting from bubble collisions and sound wave propagation in the plasma is also within the reach of future GWs observatories, such as the Einstein Telescope (ET) \cite{ET}, the Laser Interferometer Space Antenna (LISA) \cite{LISA} or the Atomic Experiment for DM and Gravity Exploration in space (AEDGE) \cite{AEDGE:2019nxb}.

Although we have implicitly assumed all the DM in the Universe to be produced by a BE mechanism, its generation might potentially occur along with other DM production channels, modifying with it the parameter space in agreement with observations. For instance, final transition temperatures exceeding the mass of scalar DM candidates coupled to the thermal bath could result in additional \textit{Freeze-Out} production, with \textit{Freeze-In} contributions becoming likely relevant in the opposite regime \cite{Azatov:2024crd}.  Also, one could consider DM production from bubble collisions \cite{Falkowski:2012fb}, although this mechanism is typically subdominant as compared to the one discussed here \cite{Azatov:2021ifm,Azatov:2024crd}. Finally, once the DM is produced, its evolution will depend on how it interacts with the thermal bath, mainly through the transition field. A priori, one could envisage a \textit{ free streaming regime} in which the scattering is out of equilibrium and the velocities of the DM particles are not modified, and a \textit{cold DM regime} in which the scattering slows down the produced particles until they reach non-relativistic velocities \cite{Azatov:2024crd}. However, the latter case is usually not relevant for DM scalar particles generated via renormalizable interactions \cite{Baldes:2022oev}, this being the output also expected for weaker interactions such as those considered in this paper. 

\section{Acknowledgments}

JL acknowledges the support of the Institute of Particle and Cosmos Physics through the ``M.Sc. grants IPARCOS-UCM/2023". JR is supported by a Ramón y Cajal contract of the Spanish Ministry of Science and Innovation with Ref.~RYC2020-028870-I. 
JARC is partially supported by the COST
(European Cooperation in Science and Technology) Actions CA22113 and CA21106. This work was supported by the project PID2022-139841NB-I00 of MICIU/AEI/10.13039/501100011033 and FEDER, UE.

\onecolumngrid

\appendix
\section{DM production}

In this appendix, we provide detailed insights into the specific calculations required to derive the density of dark matter produced via the BE mechanism in the $\tilde{\gamma}_w>1$ regime. For additional details on the renormalizable case, please refer to Appendix A of Ref.~\cite{Azatov:2021ifm}.

\subsection{DM production with 4-dimensional interaction}\label{A0}

We start with the renormalizable case ($n=2$). To begin, we derive the matrix element shown in Eq.\,(\ref{matriz M n2}) from Eq.\,(\ref{matriz M})

\begin{equation}
    \begin{split}
        \mathcal{M}=&\int_{-\infty}^{\infty} dz \,e^{i\Delta p_z z}\lambda\langle\Phi\rangle(z)=\int_0^{L_w}\mathrm{d}ze^{i\Delta p_zz}\lambda v\frac{z}{L_w}+\int_{L_w}^{\infty}\mathrm{d}ze^{i\Delta p_zz}\lambda v\\
        =&\lambda \frac{v}{L_w}\left[\frac{z}{i\Delta p_z}e^{i\Delta P_zz}+\frac{e^{i\Delta p_zz}}{\Delta p_z^2}\right]_0^{L_w}+\frac{\lambda v}{i\Delta p_z}e^{i\Delta p_zz}\Big|_{L_w}^{\infty}\\
        =&\lambda \left(\frac{v}{\Delta p_z}\right)\frac{e^{2i\sigma}-1}{2\sigma}\,,  \quad \text{being}\quad\sigma=\frac{L_w\Delta p_z}{2}\,.
    \end{split}
\end{equation}

\noindent Then, the squared matrix element is easily derive, being

\begin{equation}
    \big|\mathcal{M}\big|^2=\lambda ^2\left(\frac{v}{\Delta p_z}\right)^2 \textrm{sinc}^2\sigma \,.
\end{equation}
Now, substituting this result into Eq.~(\ref{probabilidad}) and taking into account the explicit form of the $0$-momentum components in Eq.~(\ref{momentos}), the decay probability in this case can be rewritten as
\begin{equation}
\begin{split}
    \mathcal{P}=\frac{1}{2p_0}\frac{\lambda^2v^2}{(2\pi)^3}\int\frac{\textit{d}^3k^{\phi_1}}{2p_0x}\frac{\textit{d}^3k^{\phi_2}}{2p_0(1-x)}\delta^2_{\perp}\delta_0\frac{\text{sinc}^2\sigma}{\Delta p_z^2}\,.
\end{split}
\end{equation}
The quadratic dependence on the transversal components together with the relations $k^{\phi_1}_{\perp}=-k^{\phi_2}_{\perp}$, and $(k^{\phi_1}_{\perp})^2=(k^{\phi_2}_{\perp})^2\equiv k^2_{\perp}$ allows us to integrate over $\textit{d}^2\mathbf{k}_{\perp}^{\phi_2}$ with the associated Dirac delta. Changing also $\textit{d}^3k^{\phi_1}$ to cylindrical coordinates, we get 
\begin{equation}
     \mathcal{P}=\frac{1}{2p_0}\frac{\lambda^2v^2}{(2\pi)^2}\int\frac{\textit{d}k_z^{\phi_1}\lvert \mathbf{k}_{\perp}^{\phi_1}\rvert \textit{d}k_{\perp}^{\phi_1}}{4p_0^2x(1-x)}\textit{d}k_z^{\phi_2}\delta_0\frac{\text{sinc}^2\sigma}{\Delta p_z^2}\,.
\end{equation}
To proceed further, we integrate in $x,k_{\perp}^2, k_0^{\phi_1}$. Introducing the Jacobian associated to this change of variables, $J= k_0^{\phi_1}p_0^2x/(2 k_z^{\phi_1} k_z^{\phi_2}k_{\perp})$,  making use of the remaining Dirac delta and considering the limit $p_0^2 (1-x)^2\gg k_{\perp}^2+M^2$ and $p_0^2x^2 \gg k_{\perp}^2+M^2$ such that $k_z^{\phi_1}k_z^{\phi_2}\simeq p_0^2x(1-x)$ and $\Delta p_z\simeq (k_{\perp}^2+M^2)/2p_0x(1-x)$, we get
\begin{equation}
    \mathcal{P}\simeq \frac{\lambda^2v^2}{16\pi^2}  \int\frac{\textit{d}x\textit{d}k_{\perp}^2 x(1-x)}{(k_{\perp}^2+M^2)^2}\,\text{sinc}^2\sigma\,\Theta(p_0-2M)\,,
\end{equation}
with the Heaviside function $\Theta(p_0-2M)$ ensuring that the $z$ component of the momenta is real. Note that the considered limit is consistent with a relativistic expansion of the bubbles and prevents any of the two decay products from
taking almost all the available energy ($x\simeq0$, $x\simeq1$). Finally, since the function $\text{sinc}^2 \sigma$ goes quickly to $0$ for $\sigma\gg 1$ and has a value close to $1$ otherwise, we impose that $\sigma \lesssim 1$, or equivalently, $p_0-M^2L_w >0$. This allows us to replace this function by a new step function $\Theta(p_0-M^2L_w)$, leading us to the final expression
\begin{equation}
    \mathcal{P}=\frac{\lambda^2}{96\pi^2}\left(\frac{ v}{M}\right)^2\Theta(p_0-2M)\Theta(p_0-M^2L_w)\,.
    \label{probabilidad n=2}
\end{equation}
Note that this equation corrects the result of \cite{Azatov:2021ifm} by a factor $1/4$, with the latter Heaviside function ensuring the regime $\Delta p_z L_w<1$.  Moreover, since $L_w\sim1/v$ \cite{Azatov:2021ifm}, the assumed hierarchy $M\gg v$ implies $M^2L_w>2M$.

Having determined the probability of decay \eqref{probabilidad n=2}, we proceed to compute the DM density and relic abundance. Expressing Eq.~(\ref{expresion general densidad}) in cylindrical coordinates we get 
\begin{equation}
    n_{\phi}=\frac{\lambda^2 }{384\pi^4 \gamma_w v_w} \left(\frac{v}{M} \right)^2 \int_0^{\infty}\textit{d}p_{\perp}^2\int_{M^2/v}^{\infty}\textit{d}p_z \,f_h(p_z,\vec{p}_{\perp}) \,,
    \label{calculando densidad}
\end{equation}
and consequently
\begin{equation}
\begin{split}
     n_{\phi}=\frac{\lambda^2T_{\text{n}}^3}{ 48\pi^4}& \left(\frac{  v}{ M}\right)^2   \,e^{-\frac{M^2}{v T_{\text{n}}}\gamma_w (1-v_w)}
     \frac{1}{4v_w\gamma_w^2}\left[\frac{(2-v_w)}{(1-v_w)^2\gamma^2_w}+\frac{M^2/(v T_{\rm n})}{(1-v_w)\gamma_w}\right]\,.
\end{split}
\label{densidadn2 completa}
\end{equation}
In the ultra-relativistic limit $v_w \rightarrow 1$, we have $\gamma_w=1/\sqrt{1-v_w^2}\gg 1$ and $\gamma_w(1-v_w)=\gamma_w-\sqrt{\gamma_w^2-1}\simeq 1/(2\gamma_w)$,  and therefore
\begin{equation}
    n_{\phi}\simeq \frac{\lambda^2  T_{\text{n}}^3}{48\pi^4}\left(\frac{v}{M} \right)^2 \, e^{-\frac{1}{2\gamma_w}\frac{M^2}{v T_{\text{n}}}}\simeq \frac{\lambda^2  T_{\text{n}}^3}{48\pi^4}\left(\frac{v}{M} \right)^2\,.
\label{densidad n2 aprox}
\end{equation}
The current DM relic abundance in this regime ($\gamma_w \gg 1$, $\gamma_w\gg M^2/vT_n$) is therefore given by
\begin{equation}
     ^2\Omega_{\text{BE}} h^2=2.7\cdot10^7 \frac{\lambda ^2}{g_{\star s}} \frac{v}{M}\frac{v}{200\,\text{GeV}}\left(\frac{T_{\text{n}}}{T_{\text{f}}}\right)^3\,.
     \label{parametro densidad n=2}
\end{equation}
This result agrees with that presented in Refs.~\cite{Azatov:2021ifm,Azatov:2024crd} for scalar decay products, exhibiting essentially also the same dependencies in the transition parameters as those found for fermions and vectors in Ref.~\cite{Azatov:2024crd}. Additionally, the complete expression for the abundance derived from Eq.\,(\ref{densidadn2 completa}) without further approximations is shown in Table\,\ref{table:expresiones completas}.

\begin{table*}[!t]
\begin{ruledtabular}
\begin{tabular}{ c c c c c}
\multicolumn{5}{c}{
$\displaystyle ^{2}\Omega_{\text{BE}}h^2=\frac{200\Upsilon}{192\pi^4}\frac{\lambda^2}{v_w\gamma_w^3g_{\star s}}\frac{T_n^3}{T_f^3}\frac{v}{M}v_{200}g(M^2/T_nv)$} \\[1.5ex] \hline 
 \multicolumn{5}{c}{
$\displaystyle ^{3}\Omega_{\text{BE}}h^2=A'\bigg[\frac{1}{(1-v_w)^4}\left(f_1(2M/T_n)+f_1(M^2/T_nv)-2f_1(2M^2/T_nv)\right)+\frac{1}{(1-v_w)^3}\big(f_2(2M/T_n)$} \\ [1.5ex]  
  \multicolumn{5}{c}{$\displaystyle+f_2(M^2/T_nv)-2f_2(2M^2/T_nv)\big)\bigg]+B'\bigg[g(2M^2/T_nv)-\frac{1}{4}g(M^2/T_nv)\bigg]$} \\[1.5ex]
 \multicolumn{5}{c}{with $\displaystyle A'=\frac{5\Upsilon\lambda^2}{21\pi^4\gamma_w^6v_wg_{\star s}}\frac{T_{\text{n}}^5}{T_{\text{f}}^3\Lambda^2}\frac{v^5}{M^5}v_{200}$ \, and \,   $\displaystyle B'=\frac{25\Upsilon\lambda^2}{18\pi^4\gamma_w^3 v_w g_{\star s}}\frac{T_{\text{n}}^3}{T_{\text{f}}^3}\frac{v^3}{M\Lambda^2}v_{200}$}\\[1.5ex] \hline 
\multicolumn{5}{c}{$\displaystyle ^{4}\Omega_{\text{BE}}h^2=A\bigg[\frac{1}{(1-v_w)^4}\left(f_1(2M/T_n)-\frac{7}{3}f_1(M^2/T_nv)+\frac{4}{3}f_1(2M^2/T_nv)\right)
+\frac{1}{(1-v_w)^3}\big(f_2(2M/T_n)$} \\[1.5ex]\multicolumn{5}{c}{ $\displaystyle-\frac{7}{3}f_2(M^2/T_nv)+\frac{4}{3}f_2(2M^2/T_nv)\big)\bigg]+B\bigg[\frac{1}{(1-v_w)^6}\big(f_3(2M/T_n)+f_3(M^2/T_nv)-2f_3(2M^2/T_nv)\big)$}   \\ [1.5ex] 
\multicolumn{5}{c}{$\displaystyle +\frac{1}{(1-v_w)^5}\left(f_4(2M/T_n)+f_4(M^2/T_nv)-2f_4(2M^2/T_nv)\right)\bigg]+C\bigg[\frac{25}{48}g(M^2/T_nv)-\frac{1}{5}g(2M^2/T_nv)\bigg]$} \\ [1.5ex]
\multicolumn{5}{c}{ with $\displaystyle A= \frac{15\Upsilon\lambda^2}{28\pi^4g_{\star s} v_w \gamma_w^6}\frac{T_{\text{n}}^5}{M^5}\frac{v^4}{\Lambda^4}\frac{v^3}{T_{\text{f}}^3}v_{200}$\,, \, $\displaystyle  B=\frac{20\Upsilon \lambda^2}{77\pi^4g_{\star s}\gamma_w^8 v_w}\frac{T_{\text{n}}^7}{\Lambda^4 T_{\text{f}}^3}  \frac{v^{9}}{M^9 } v_{200}$ \, and \, $\displaystyle C=\frac{25\Upsilon\lambda^2}{8\pi^4\gamma_w^3v_w g_{\star s}}\frac{v^5T_n^3}{M\Lambda^4T_f^3}v_{\text{200}}$}  \\ \hline \hline
 \multicolumn{2}{c}{ } &\\ [0.1ex] 
 \multicolumn{5}{c}{$\displaystyle f_n(x)=\left[\sum_{i=0}^j{\left(\gamma_w(1-v_w)x\right)^i\frac{j!}{i!}}\right]e^{-\gamma_w(1-v_w)x};\quad g(x)=e^{-\gamma_w x (1-v_w)}\left[\frac{(2-v_w)}{(1-v_w)^2\gamma_w}+\frac{x}{(1-v_w)}\right]$ } \\ [1.5ex] 
 \multicolumn{5}{c}{$\displaystyle \text{with } j=3 \text{ for } n=1 \text{, } j=2 \text{ for } n=2 \text{, } j=5 \text{ for } n=3 \text{ and } j=4 \text{ for } n=4.$}   \\ [1.5ex] 
\end{tabular}
\end{ruledtabular}
\caption{\label{table:expresiones completas} Complete expressions of the DM relic abundance for $n=2$, $n=3$ and $n=4$. The quantity $v_{200}$ stands for the vev on the non-symmetric side of the wall normalized by $200$ GeV, $v_{200}=v/(200\,\text{GeV})$.}
\end{table*}

\subsection{DM production with 5-dimensional interaction}\label{A1}
Now, we present the details of the computation for the 5-dimensional case ($n=3$), following very similar steps.  As a first step, we derive Eq. (\ref{M n3}) from Eq. (\ref{matriz M})
\begin{equation}
\begin{split}
    \mathcal{M}=&\int_{-\infty}^{\infty} dz \,e^{i\Delta p_z z}\frac{\lambda}{\Lambda}\left(\langle\Phi\rangle(z)\right)^2=\int_0^{L_w}dz e^{i\Delta p_z z} \frac{\lambda}{\Lambda} \frac{v^2z^2}{L_w^2}+\int_{L_w}^{\infty}dz e^{i\Delta p_z z} \frac{\lambda}{\Lambda}v^2\\=&\frac{\lambda v^2}{\Lambda L_w^2}\left[\left(\frac{z^2}{i\Delta p_z}-\frac{2z}{(i\Delta p_z)^2}+\frac{2}{(i\Delta p_z)^3}\right)e^{i\Delta p_z z}\right]^{L_w}_0+\frac{\lambda v^2}{\Lambda }\left.\frac{e^{i\Delta p_z z}}{i \Delta p_z}\right|_{L_w}^{\infty}\\=&\frac{2\lambda v^2}{\Lambda (i\Delta p_z)^3L_w^2}\left[e^{i\Delta p_z L_w}\left(1-i\Delta p_z L_w\right)-1\right]=2i\lambda\left(\frac{v}{\Delta p_z}\right)^2\frac{1}{L_w\Lambda}\left[\frac{e^{2i\sigma}-1}{2\sigma}-ie^{2i\sigma} \right]\,,
\end{split}
\end{equation}
and with it the squared matrix element
\begin{equation}\hspace{-3mm}
\big|\mathcal{M} \big|^2=4\lambda^2 \left( \frac{v}{\Delta p_z} \right)^4 \left(\frac{1}{\Lambda L_w}\right)^2\left(\text{sinc}^2\sigma+1-2\,  \text{sinc\,}2\sigma\right)\,.
\label{amplitud cuadrado n3}
\end{equation}
Considering the frame presented in Section \ref{BE mechanism} and the kinematics in Eq.~(\ref{momentos}), we obtain can obtain the probability via Eq.\,(\ref{probabilidad}):
\begin{equation}
\begin{split}
    \mathcal{P}=\frac{2}{p_0(2\pi)^3}\left(\frac{\lambda v^2}{\Lambda L_w}\right)^2\int\frac{\textit{d}^3k^{\phi_1}}{2p_0x}\frac{\textit{d}^3k^{\phi_2}}{2p_0(1-x)}\delta^2_{\perp}\delta_0\frac{1}{\Delta p_z^4}\left(\text{sinc}^2\sigma+1-2\,  \text{sinc\,}2\sigma\right)\,,
\end{split}
\end{equation}
where we have taken into account the explicit form of the 0-momentum components.  Now, since $(k_{\perp}^{\phi_1})^2=(k_{\perp}^{\phi_2})^2$ we can use the transversal Dirac delta to integrate over $\textit{d}^2k_{\perp}^{\phi_2}$. Expressing $\textit{d}^3k_{\perp}^{\phi_1}$ in cylindrical coordinates and replacing $ \Delta p_z\simeq (k_{\perp}^2+M^2)/2p_0x(1-x)$, we get
\begin{equation}
\begin{split}
\mathcal{P}= \frac{8}{p_0(2\pi)^2}\left(\frac{\lambda v^2}{\Lambda L_w}\right)^2\int \textit{d}k_z^{\phi_1}\lvert \mathbf{k}_{\perp}^{\phi_1}\rvert \textit{d}k_{\perp}^{\phi_1}\textit{d}k_z^{\phi_2}\delta_0\frac{x^3(1-x)^3p_0^2}{(M^2+k_{\perp}^2)^4}\left(\text{sinc}^2\sigma+1-2\text{sinc}\,2\sigma\right)\,.
\end{split}
\end{equation}
  To proceed further,  we change the variables $k_z^{\phi_1},\,k_{\perp}^{\phi_1},\,k_z^{\phi_2}$ to $x,\,k_{\perp}^2,\, k_0^{\phi_1}$, introduce the associated Jacobian  $J= k_0^{\phi_1}p_0^2x/(2 k_z^{\phi_1} k_z^{\phi_2}k_{\perp})$ and integrate over $\textit{d}k_0^{\phi_1}$ using the remaining Dirac delta, being the only dependence on this variable the one introduced by the Jacobian. Taking now the limits $p_0^2 (1-x)^2\gg k_{\perp}^2+M^2$ and $p_0^2x^2 \gg k_{\perp}^2+M^2$ we can approximate $k_z^{\phi_1}k_z^{\phi_2}\simeq p_0^2x(1-x)$, obtaining with it
\begin{equation}
    \begin{split}
 \mathcal{P}= \frac{4}{p_0(2\pi)^2}\left(\frac{\lambda v^2}{\Lambda L_w}\right)^2\int \textit{d}x\textit{d}k_{\perp}^2 \frac{x^3(1-x)^3p_0^3}{(M^2+k_{\perp}^2)^4}\left(\text{sinc}^2\sigma+1-2\text{sinc}\,2\sigma\right)\Theta(p_0-2M)\,.
    \end{split}
\end{equation}
Last, to be able to compute this final integral, we replace the $\text{sinc}\,x$ functions  by the first terms of their Taylor's series, i.e. $\text{sinc} x\simeq 1-x^2/6+x^4/120$, and introduce Heaviside functions ensuring respectively that $\sigma<1$ and $2\sigma<1$.   Namely, we have taken all the terms of the expansion that leave us with positive powers of $p_0$ (two or three terms depending of the case).  This way, in this regime, the sought probability takes the form 
\begin{equation}
\begin{split}
    \mathcal{P}=\frac{\lambda^2}{24 \pi^2}  \left(\frac{v}{M}\right)^4 \left( \frac{1}{\Lambda L_w}\right)^2
     \Theta(a)\Theta(b) \bigg[\frac{2}{35} \left(\frac{p_0}{M}\right)^2\times\big(1+\Theta(c)-2\Theta(d)\big)
        +\frac{\left(M L_w\right)^2}{3}\bigg(\Theta(d)-\frac{1}{4}\Theta(c)\bigg)\bigg]\,,
\end{split}
\end{equation}
where $a=(p_0-2M)$, $b=(\Lambda^2-2p_0v)$, $c=(p_0-M^2L_w)$ and $d=(p_0-2M^2L_w)$. Moreover, the Heaviside function $\Theta (b)$ ensures that the effective field theory is valid, guaranteeing that the energy in the center of mass of the particle-bubble system is lower than $\Lambda$. To calculate the associated DM density, it is important to recall that $p_0=p_0(p_z)$. For a relativistic expansion of the bubbles, $p_0\simeq p_z$, we have
 \begin{equation}
     n_{\phi}\simeq \frac{2}{\gamma_w v_w}\int \frac{\textit{d}^3p}{(2\pi)^3} \, \mathcal{P} \, \exp\left(-\frac{\gamma_w(\sqrt{p_z^2+p_{\perp}^2}-v_w p_z)}{T_{\text{n}}}\right)\,.
 \end{equation}
Since the probability does not depend on $p_{\perp}$, the easiest way to proceed is to use cylindrical coordinates. Changing $\textit{d} p_{\perp} p_{\perp}$ to $\textit{d}p_{\perp}^2$, we obtain
 \begin{equation}
 \begin{split}
     n_{\phi}\simeq \frac{2\pi}{\gamma_w v_w}\int_{-\infty}^{\infty} \frac{\textit{d}p_z}{(2\pi)^3} \, \mathcal{P}(p_z) \, e^{\frac{\gamma_w}{T_n}v_wp_z}\int_0^{\infty} \textit{d}p_{\perp}^2 e^{\frac{-\gamma_w\sqrt{p_z^2+p_{\perp}^2}}{T_{\text{n}}}}=\frac{2\pi}{\gamma_w v_w}\int_{-\infty}^{\infty} \frac{\textit{d}p_z}{(2\pi)^3} \, \mathcal{P}(p_z) \, e^{\frac{\gamma_w}{T_n}v_wp_z}\cdot2e^{-\frac{\gamma_w}{T_n}p_z}\frac{\frac{\gamma_w p_z}{T_n}+1}{(\frac{\gamma_w}{T_n})^2}\,.
 \end{split}
      \label{densidad anexo}
 \end{equation}
The final integral involves summing integrals of the form $\int dx \, x^m \exp(-\gamma_w x (1-v_w)/T_n)$, leading to the results shown in Table \ref{table:expresiones completas}.

\subsection{DM production with 6-dimensional interaction}\label{A2}
While the procedure for $n=4$ is completely analogous to that for $n=3$, we present here some of the intermediate results for the sake of completeness. From the matrix element 
\begin{equation}
\begin{split}
     \mathcal{M}&=\int_{-\infty}^{\infty} dz e^{i\Delta p_z z}\frac{\lambda}{\Lambda^2}\left(\langle\Phi\rangle(z)\right)^3=\left.\frac{\lambda v^3}{i\Lambda^2\Delta p_z}e^{i\Delta p_z z}\right|_{L_w}^{\infty}+\frac{\lambda v^3}{\Lambda^2 L_w^3}\bigg[e^{i\Delta p_z z}\bigg(\frac{z^3}{i\Delta p_z}-\frac{3z^2}{(i\Delta p_z)^2}+\frac{6z}{(i\Delta p_z)^3}-\frac{6}{(i\Delta p_z)^4}\bigg)\bigg]_0^{L_w}\\ &=6\lambda \left(\frac{v}{\Delta p_z}\right)^3\left(\frac{1}{\Lambda L_w}\right)^2 \left[\frac{1-e^{2i\sigma}}{2\sigma}+e^{2i\sigma}\left(\sigma+i\right)\right]\,,
\end{split}
\end{equation}
we obtain 
\begin{equation}
\begin{split}
    \big|\mathcal{M}\big|^2=36\lambda^2 \left(\frac{v}{ \Delta p_z}\right)^6\left(\frac{1}{\Lambda  L_w}\right)^4 \, \left[ \text{sinc}^2\sigma+1-2 \,\text{sinc}\, 2 \sigma+\sigma ^2(1-2
   \text{sinc}^2\sigma)\right]\,.
\end{split}
\end{equation}
Once again, the powers of $1/\Delta p_z$ increase with the dimension of the operator. With the same changes of variables and limits as before, the corresponding probability (Eq.\,\eqref{probabilidad}) takes the form 
\begin{equation}
\mathcal{P}= \frac
         {18\lambda^2}{(2\pi)^2}\frac{v^6}{p_0(\Lambda L_w)^4}\int\textit{d}x\textit{d}k_{\perp}^2\Theta(p_0-2M)\frac{p_0^3(1-x)^3x^3}{(M^2+k_{\perp}^2)^4} \left[\frac{L_w^2}{2}(1-2\text{sinc}^2\sigma)+8\left(\frac{p_0(1-x)x}{M^2+k_{\perp}^2}\right)^2(\text{sinc}^2\sigma+1-2\text{sinc}2\sigma)\right]\,.
\end{equation}
Expanding again the $\text{sinc}^2\sigma$ and $\text{sinc}\,2\sigma$ functions and introducing the corresponding Heaviside function to ensure that these approximations remain valid, we obtain
\begin{equation}
\begin{split}
     \mathcal{P}=\frac{3\lambda^2}{8\pi^2}\left(\frac{v}{M}\right)^6 \left(\frac{1}{\Lambda  L_w}\right)^4 \Theta(a)\Theta(b) \bigg[\frac{8}{1155}\left(\frac{p_0}{M}\right)^4\left(1+\Theta(c)-2\Theta(d)\right)\\+\frac{(p_0 L_w)^2}{70} \bigg(1-\frac{7}{3}\Theta(c) +\frac{4}{3}\Theta(d)\bigg)
        +\frac{1}{12}(L_wM)^4\left(\frac{25}{48}\Theta(c)-\frac{1}{5}\Theta(d)\right)
    \bigg]\,.
\end{split}
\end{equation}
The discussion regarding $n_{\phi}$ in the previous subsection remains applicable, as well as the expression in Eq.~(\ref{densidad anexo}), leading us to the results presented in Table\,\ref{table:expresiones completas}.

\twocolumngrid
\bibliography{biblio.bib}

\end{document}